\documentclass[3p,times]{elsarticle}

\begin{document}

\begin{frontmatter}

\title{Reliability of COVID-19 data and government policies.}

\author{T.~M.~Rocha Filho\corref{cor1}}
\ead{marciano@fis.unb.br}
\address{International Center for Condensed Matter Physics and Instituto de Física,\\ Universidade de Brasília, Brasília -- BRAZIL.}
\author{J.~F.~F.~Mendes}
\address{Departamento de Física and I3N, Universidade de Aveiro, 3880 Aveiro -- Portugal}
\author{M.~L.~Lucio}
\address{Departamento de Gestão de Políticas Públicas/FACE, Universidade de Brasília, Brasília -- BRAZIL.}
\author{M.~A.~Moret}
\address{Centro Universitário SENAI CIMATEC and \\ Universidade do Estado da Bahia, Salvador -- Brazil}

\begin{abstract}
We study how available data on COVID-19 cases and deaths in different countries are reliable. Our analysis is based
on a modification of the law of anomalous numbers, the Newcomb-Benford law, applied to the daily number of
deaths and new cases in each country. We first revisit the Newcomb-Benford law and show how to avoid false negative
compliance of the data. We then compared deviation from this law, to a number of social and
economic indices for each country by computing the Spearman rank order correlation between each index and the $\chi^2$
deviation of COVID-19 data to the modified NB law.
A similar analysis for excess deaths for the same countries with sufficient available data was performed.
We conclude that in general less democratic, less transparent and more corrupt countries
tend to have data of lesser quality. We also discuss the limitations of the present approach.
\end{abstract}


\end{frontmatter}

\section{Introduction}

The still ongoing COVID-19 pandemic has caused in the whole world, from official data, close to 600 million cases
and $6.4$ million deaths~\cite{jh}.
Underreporting is a known issue, with the real number of cases estimated
to vary from 3.2 to 18 times the official count across different countries~\cite{rahmandad},
while deaths are underreported to a lesser extend~\cite{adam,economist}.
Different explanations can be brought to front.
Insufficient testing: up to the present day, Denmark and USA performed 21.8 million and 3 million tests per million inhabitants,
respectively, while Brazil 296 thousands and China 111 thousands only~\cite{worldometer}.
A deficient health system: many low and middle income countries lack proper health care under normal circumstances,
a situation which is only worsen by the pandemic.
And last but not least,
the unwillingness of some authorities to properly
account for the situation, or even plainly denying the existence of the pandemic~\cite{malta}.
Consequently, the reliance on COVID-19 official data must be carefully weighted for each locality.

The quality and reliability of COVID-19 data is essential for a proper planning and mitigation of the COVID-19
pandemic, and had an important impact in the quality of different studies~\cite{ronquillo,rajan,jung,starnini,stoto}.
Knutsen and Kolvani~\cite{knutsen}
discussed regime type and state capacity with relation to the capability to respond to COVID-19,
how autocracies tend to under-report COVID-19 deaths, and the positive relationship between democracy level, state capacity
and  the ability to mitigate the pandemic. Annaka~\cite{annaka} studied the
relationship between data transparency for different political regimes and COVID-19 deaths, and concluded that
transparency is positively correlated with the number of death cases more consistently, pointing to data
manipulation by autocratic regimes.

Some works have been devoted to study data quality and the possibility of plain manipulation by different governments
using forensic methods, with a preference for the Newcomb-Benford (NB) law~\cite{nigrini},
an empirical law that predict the frequency
of the first significant number in many naturally occurring data (see next section), and was used to identify anomalous data
in the series of cases and deaths for different countries or in specific locations. Neumayer and Plümper~\cite{neumayer}
used this approach to discuss the so-called autocratic advantage in the mitigation of the COVID-19 pandemic, when autocratic
regimes tend to have a smaller fatality ration than democratic regimes. They concluded that in fact such regimes tend to
manipulate data more frequently, thus the apparent advantage. Farhadi~\cite{farhadi} considered data for 202 countries up to September 2020,
and found that only six (Islamic Republic of Iran, El Salvador, Latvia,
São Tome and Principe, Taiwan, and Tajikistan) didn't comply with the NB law, with the other
countries fully complying or complying partially. He also showed that countries with a higher Global Health Security Index~\cite{ghsindex}
tend to comply better with the NB law. Kennedy and Yam~\cite{kennedy} discussed the conformity of COVID-19 data to
the NB law only for the initial stages of the pandemic, considering total cumulative figures, and obtained and overall
reasonable conformity. Farhadi and Lahooti~\cite{lahooti} considered two years of COVID-19 data for 198 countries,
and obtained very different agreement to NB law among them. Balashov, Yan and Zhu~\cite{balashov}
studied the relationship between results for deviations from the NB law and countries stage of development,
and compared them to economic indices. They concluded that democratic regimes and economically developed countries
tend to have more reliable data. Silva and Figueiredo Filho~\cite{silva} considered the number of cases
and deaths in Brazil, for the period from February 25 to September 15, 2020, and concluded the data
do not comply with the NB law. A similar result for essentially the same period of time was obtained by
Galvêas, Barros and Fuzo~\cite{galveas}. Both works analyzed the total number of cases and deaths, and not the daily
number which has more pronounced statistical fluctuations and thus a higher tendency to follow the NB law (see below).
Kolias~\cite{kolias} considered daily cases and deaths in the European Union for the period from March to December 2021,
with results ranging from good conformity to non-conformity among the countries,
and, quite surprisingly, that data for countries with a higher vaccination rate tend to deviate from the NB law.

In the present paper we first discuss in Section~\ref{secrevisit} a revisited implementation of the NB law, based on a
rescaling of the data to be analyzed. This results in a more consist test, with less false negative results.
Data sources are specified in Sec.~\ref{datasourcesec}.
Our approach is then applied in Sec.~\ref{secresults} to all countries in the World with at least one thousand deaths by April, 29 2022
(129 countries). The deviation from the NB law is then compared to a number of relevant economic and social indices.
Complementary to the NB law test, we also discuss the relation between the estimated proportion of excess deaths
by COVID-19 in each country, a proxy for the quality of the mitigation policies, to the same social and economic indices.
We analyze also data at the regional and city level for the USA and Brazil, the two countries with the highest official
number of deaths. We discuss our results and draw some conclusion in Section.~\ref{secdiss}.

\section{Revisiting the Newcomb-Benford law}
\label{secrevisit}

The law of anomalous numbers, or Newcomb-Benford law, states that in many naturally occurring sets of data the
frequency $P(d_1)$ of the first significant digit $d_1$ is not uniform, but follows the   distribution:
\begin{equation}
	\label{NBlaw}
	P(d_1)=\log\left(1+\frac{1}{d_1}\right),
\end{equation}
where the logarithm is computed in base $10$ (the sane law is observed in different bases~\cite{berger}).
A similar expression was also proposed for the first two significant
digits $d_1$ and $d_2$ as:
\begin{equation}
        \label{NBlaw2}
        P(d_1d_2)=\log\left(1+\frac{1}{d_1d_2}\right).
\end{equation}
The law was first discovered by Newcomb~\cite{newcomb} in 1881 and later rediscovered by Benford in 1938~\cite{benford}.
Examples of data in accordance (or significantly close to) with the distributions in Eqs.~(\ref{NBlaw}) and~(\ref{NBlaw2})
include pulsar data~\cite{shao}, astrophysical sources~\cite{moretastro},
physical properties of exoplanets~\cite{shukla}, structure of proteins~\cite{moret},
geophysical parameters~\cite{sambridge}, and fraud detection in accounting data~\cite{nigrini}, among many other applications.
We will restrict our discussions here to the first digit law only, but a similar analysis
is also possible for the more general case with second digits or further.

One important aspect of the distributions in Eqs.~(\ref{NBlaw}) and~(\ref{NBlaw2}) is that it was first formulated
as an empirical distribution, derived from raw data and insights, and not from some first principles or mathematical
properties of a field of numbers.
Different approaches were used to explain why the NB law occurs in data of so different origins. For instance, Hill~\cite{hill}
showed that the significant digits frequency of random samples from randomly selected distributions
converge to the NB law. More relevant to the present work, Rodriguez showed that random numbers sampled from a log-normal distribution with
a high variance obey closely the NB law, and that data resulting from the product of two different sets is closer to the NB
law than the original sets~\cite{rodriguez}. Log-normal distributions are common in nature and result from a central limit theorem
for multiplicative random processes, occurring for instance in disease dynamics where the effects are a result of different random
factors. The Log-normal distribution function is given by
\begin{equation}
	f(x)=\frac{1}{\sqrt{2\pi}\,\sigma x}e^{-(\ln(x)-\mu)^2/2\sigma^2},
	\label{lognormdist}
\end{equation}
where the parameters $\mu$ and $\sigma$ are the average and standard deviations of $\ln(x)$, respectively.

\begin{figure}[ht]
        \begin{center}
               {\includegraphics[width=120mm]{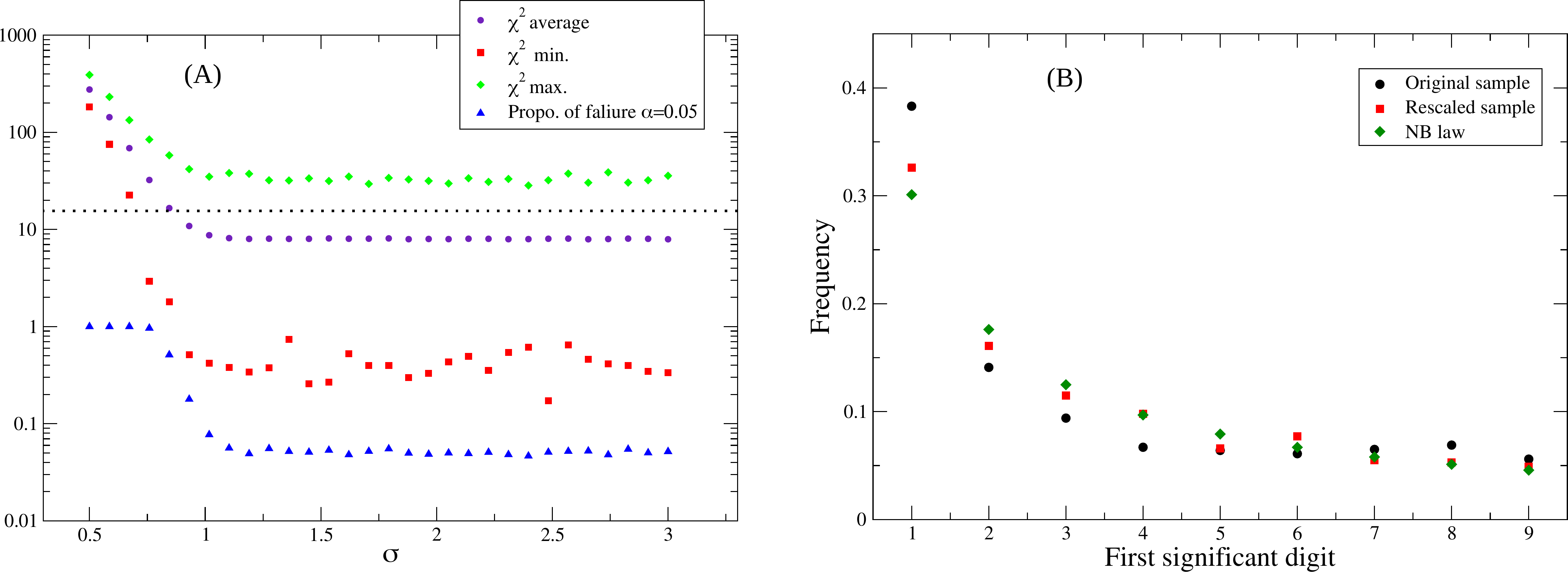}}
       \end{center}
	\caption{(A) $\chi^2$ statistic test for a sample of 1000 random numbers drawn from a log-normal distribution
	with $\mu=0$ as a function of $\sigma$ averaged over $10^4$ realizations,
	the corresponding minimal and maximal values of $\chi^2$ and the proportion of realizations that failed the test
	for significance $\alpha=0.05$. The doted line is the $\chi^2$ reference critical value of $15.507$ for significance
	$\alpha=0.05$ and 8 degrees of freedom (9 possible first significant digit minus one).
        (B) First significant digit frequency for a sample of 1000 random numbers drawn from a log-normal
        distribution with $\mu=0.1$ and $\sigma=0.7$ (circles) and for the rescaled data with zero average and
        $\sigma^\prime=1.2$ (squares). The Newcomb-Benford law is draw for comparison purposes (diamonds). The $\chi^2$
        statistic values are $\chi^2=59.01$ for the original data and $\chi^2=8.32$ for the rescaled data.
	\label{figChi2}}
\end{figure}

Figure~\ref{figChi2}A shows the average value of of the $\chi^2$ statistic
for $1000$ random numbers drawn from a log-normal distribution, with $\mu=0$,
and ten thousand realizations, as a function of $\sigma$. For a significance level of $\alpha=0.05$ and $\sigma\gtrsim 1.5$
the proportion of realizations that fail the test to reject the null hypothesis (that the random numbers do not obey the NB law)
is roughly 5\%. As discussed by Rodriguez~\cite{rodriguez}, most of natural instances of the NB law are of data that roughly or
closely follow a log-normal distribution. This is the case, for instance, for the examples discussed in the previous paragraph.
If a given sample is derived from a log-normal distribution with $\sigma<1$ it will not obey the NB law, despite the fact that
no data manipulation occurred. In order to put different samples derived for various values of $\sigma$ in the same ground,
it is reasonable to rescale the logarithms of the data such that all have $\mu=0$ and the same $\sigma$, with a sensible value
of $\sigma=1.2$ (not too much higher values can also be used).
As an example we generated a sample of 1000 random numbers drawn from a log-normal distribution
with $\mu=0.1$ and $\sigma=0.7$, with the corresponding first significant digit frequency shown in Fig.~\ref{figChi2}B.
We then rescale the data as:
\begin{equation}
\ln x^\prime=(\ln x-\mu)\times \sigma^\prime/\sigma,
	\label{resc_def}
\end{equation}
where $x$ denotes the original value of a point data, $x^\prime$ the corresponding rescaled value, $\mu$ and $\sigma$
the average and standard deviation of the original data. This rescaling is such that the rescaled random numbers
now have a log-normal distribution with zero average and standard deviation $\sigma^\prime$, chosen from the above
discussion as $\sigma^\prime=1.2$. The first significant digits distribution is shown in Fig.~\ref{figChi2}B.
We see that although the original data do not conform to the NB law, the rescaled data does.

\begin{figure}[ht]
        \begin{center}
               {\includegraphics[width=150mm]{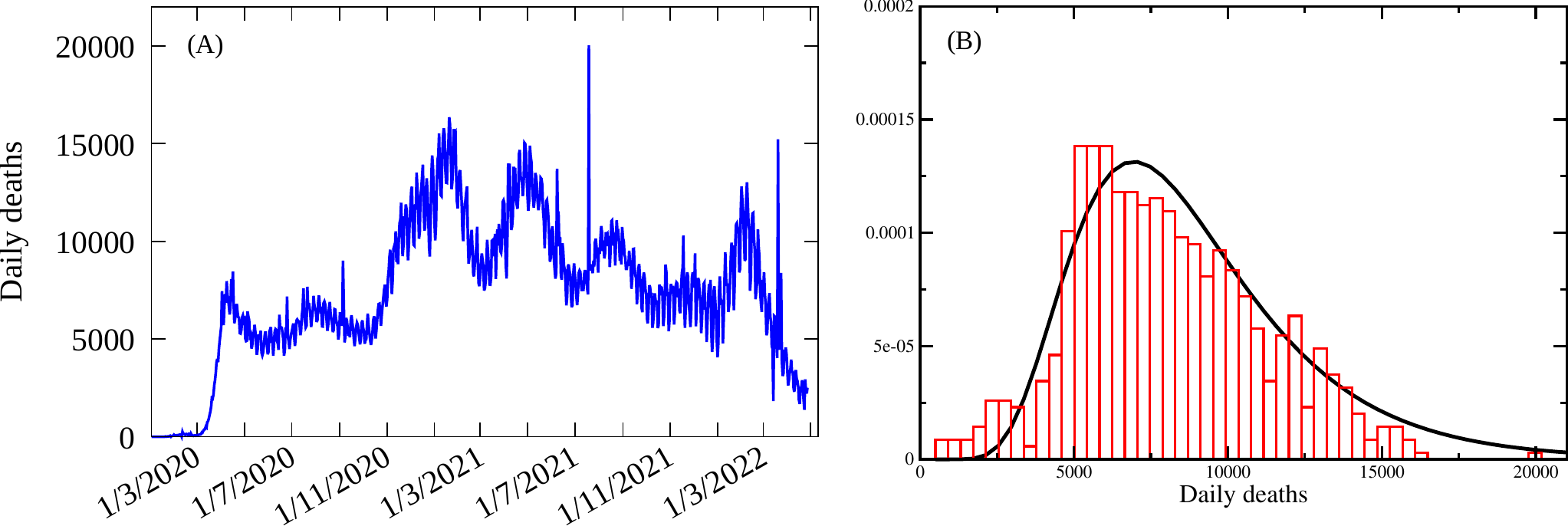}}
       \end{center}
	\caption{(A) Daily deaths by COVID-19 in the world. (B) Normalized histogram for the frequency of daily deaths
	starting at the first day with 500 deaths. The continuous line is the best fit for a log-normal distribution,
	with parameters $\mu=9.007$ ans $\sigma=0.404$\label{world_daily}}
\end{figure}

\begin{figure}[ht]
        \begin{center}
               {\includegraphics[width=65mm]{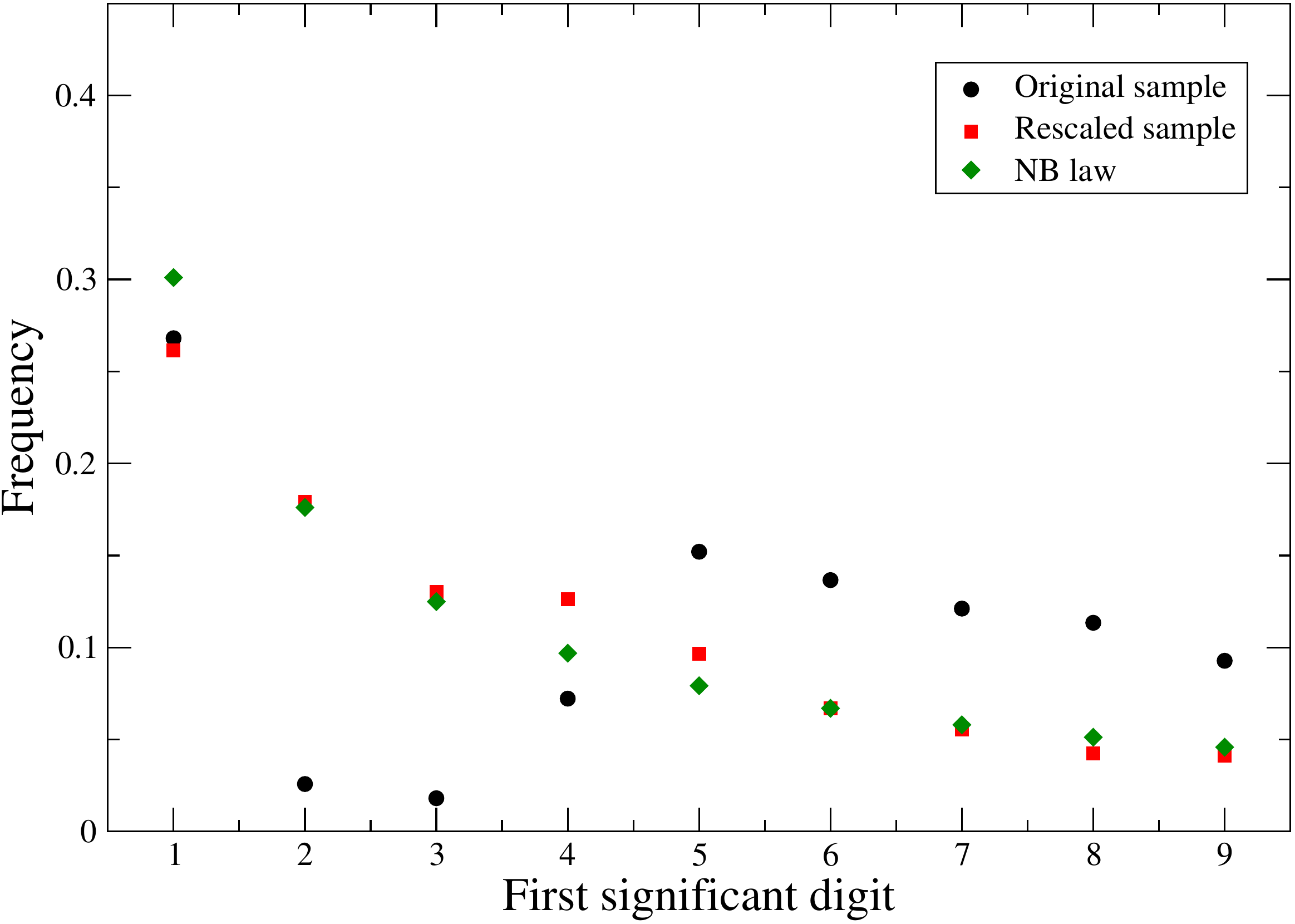}}
       \end{center}
        \caption{First significant digit frequency for the daily deaths by COVID-19 in the world for the original data
	and the rescaled data with $\sigma^\prime=3$ and $\mu^\prime=0$.\label{world_nb}}
\end{figure}

The number of daily cases during an epidemic
can be expressed, at least in some simplified epidemiological models, as proportional
to the number of susceptible and infected individuals in a population, with the proportionality factor
dependent on the structure of social contacts,
behavioral factors such as isolation and other non-pharmaceutical interventions~\cite{nos1}. All these fluctuate
randomly around mean values and thus results in a complicated multiplicative stochastic process~\cite{black}.
The same reasoning applies for the number of daily deaths.
Therefore, it is expected that the data for daily cases and deaths during the epidemic follows, at least roughly, a log-normal distribution.
In order to give some evidence for this statement, let us now consider the current COVID-19 pandemic by considering the daily number
of deaths by COVID-19 in the world. The corresponding data is available at the World Health Organization (see section~\ref{datasourcesec} below).
The number of daily deaths by COVID-19 in the whole world is shown in Fig.~\ref{world_daily}A, and the normalized histogram for the data
after the first day with $500$ deaths (to avoid too noisy data in the beginning of the pandemic) in Fig.~\ref{world_daily}B, 
with the best fit of a log-normal distribution. The first digit frequency obtained from this data is shown in Fig.~\ref{world_nb},
with a $\chi^2$ statistic of $436.183$, which is much greater than the critical value of $15.507$
with significance level $\alpha=0.05$. This is clearly due to the small value of $\sigma=0.404$, despite
the data being reasonably closer to a log-normal distribution. By rescaling the data as explained above for a higher value
of$\sigma^\prime=3$, we obtain the distribution
shown in Fig.~\ref{world_nb}, with $\chi^2=15.684$, and thus complies with the NB law.

\begin{figure}[ht]
\begin{center}
	{\includegraphics[width=140mm]{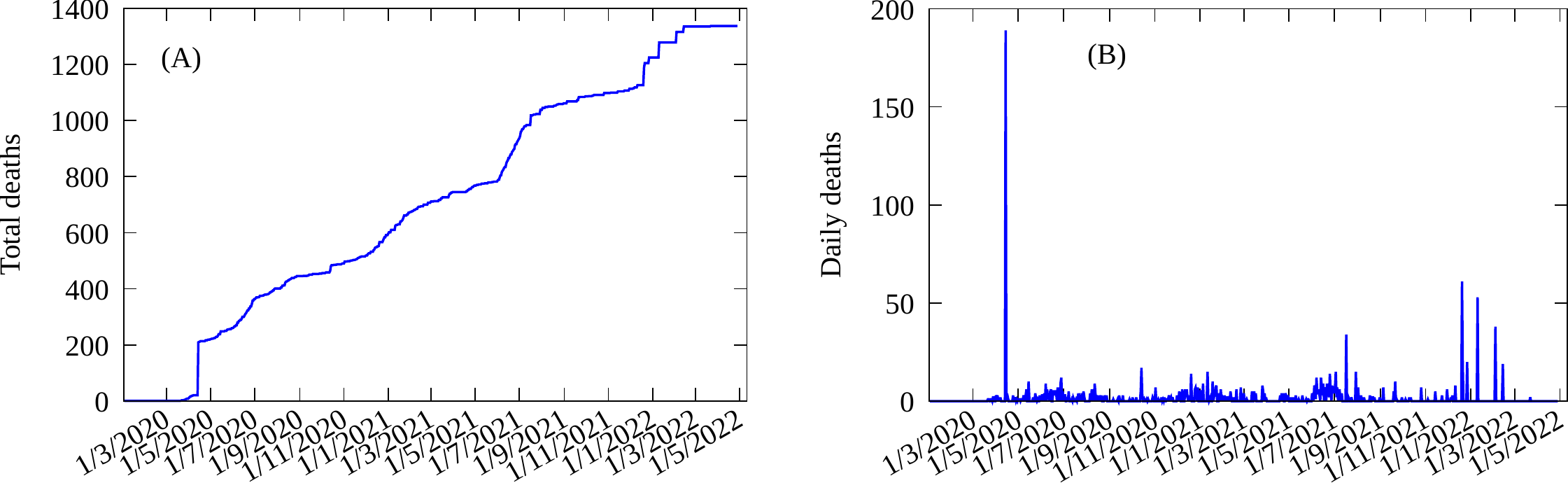}}
\end{center}
	\caption{(A) COVID-19 total deaths in the Democratic Republic of Congo.
	(B) Daily deaths.\label{excongo}}
\end{figure}

\begin{figure}[ht]
\begin{center}
        {\includegraphics[width=140mm]{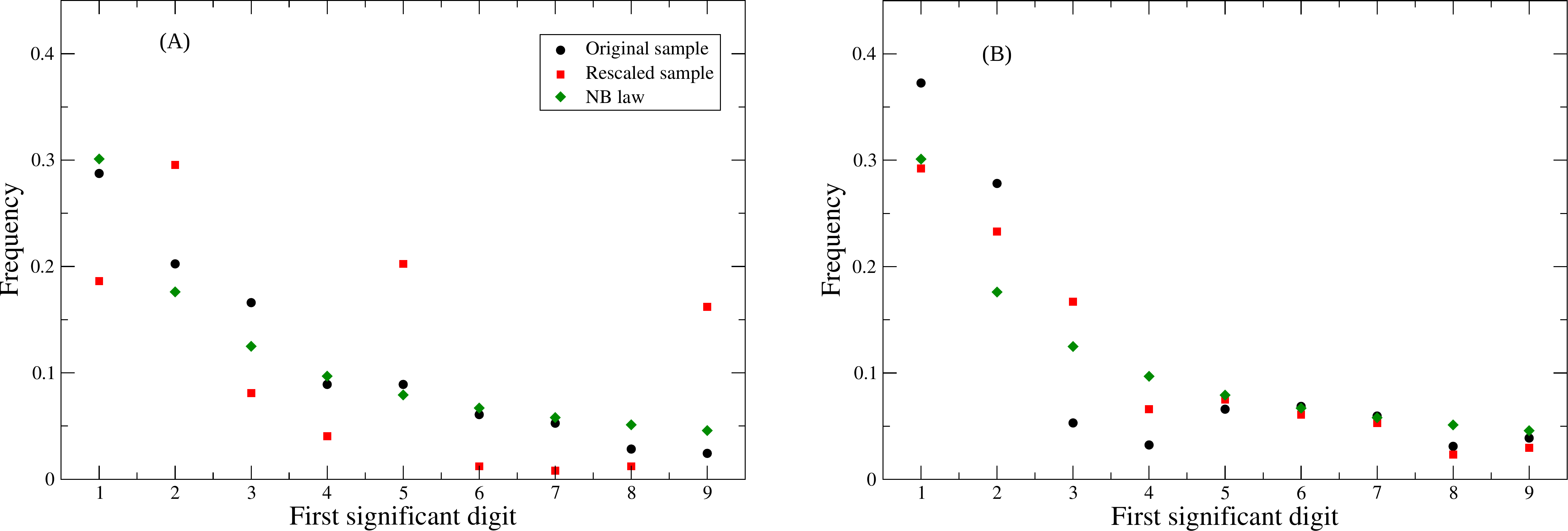}}
\end{center}
	\caption{(A) First significant digit for the number of daily deaths in the Democratic Republic
	of Congo for the original and rescaled data. (B) Same as (A) but for Turkey.\label{excongo_nb}}
\end{figure}

Let us also consider the data for daily deaths for two countries: the Democratic Republic of Congo
and Turkey. For the former data for total and daily deaths are shown in Fig.~\ref{excongo}. The frequency of data in Fig.~\ref{excongo}B
is close to a log-normal distribution with $\chi^2=18.716$ for a critical value of $66.339$ corresponding to a $50$ bins histogram. Nevertheless,
a close inspection of the data points to some
anomalies in the data, with a too high number of deaths in a few days, possibly corrections in non previously notified deaths.
Despite that, we see from Fig.~\ref{excongo_nb} that the data for daily deaths obey the NB law, with $\chi^2=10.183$.
Despite that, the rescaled data identifies straightforwardly the anomalies, with a large deviation
from the NB law with a $\chi^2=192.048$. For Turkey the series of daily deaths do not conform
to NB law, with $\chi^2=132.781$, but the rescaled data is much closer to conformity, with $\chi^2=49.844$, as shown in
Fig.~\ref{excongo_nb}. In fact the original data for Turkey is close to a log-normal distribution, with parameters
$\mu=4.51$ and $\sigma=0.95$ with $\chi^2=49.351$ for a $50$ bins histogram. Therefore, despite the fact that
the original data for daily deaths in Turkey being close to a log-normal distribution,
the NB law test would point to a larger deviation. The rescaled data is
closer to comply wit the NB law, as expected for data following roughly a log-normal distribution.

\begin{figure}[ht]
\begin{center}
        {\includegraphics[width=70mm]{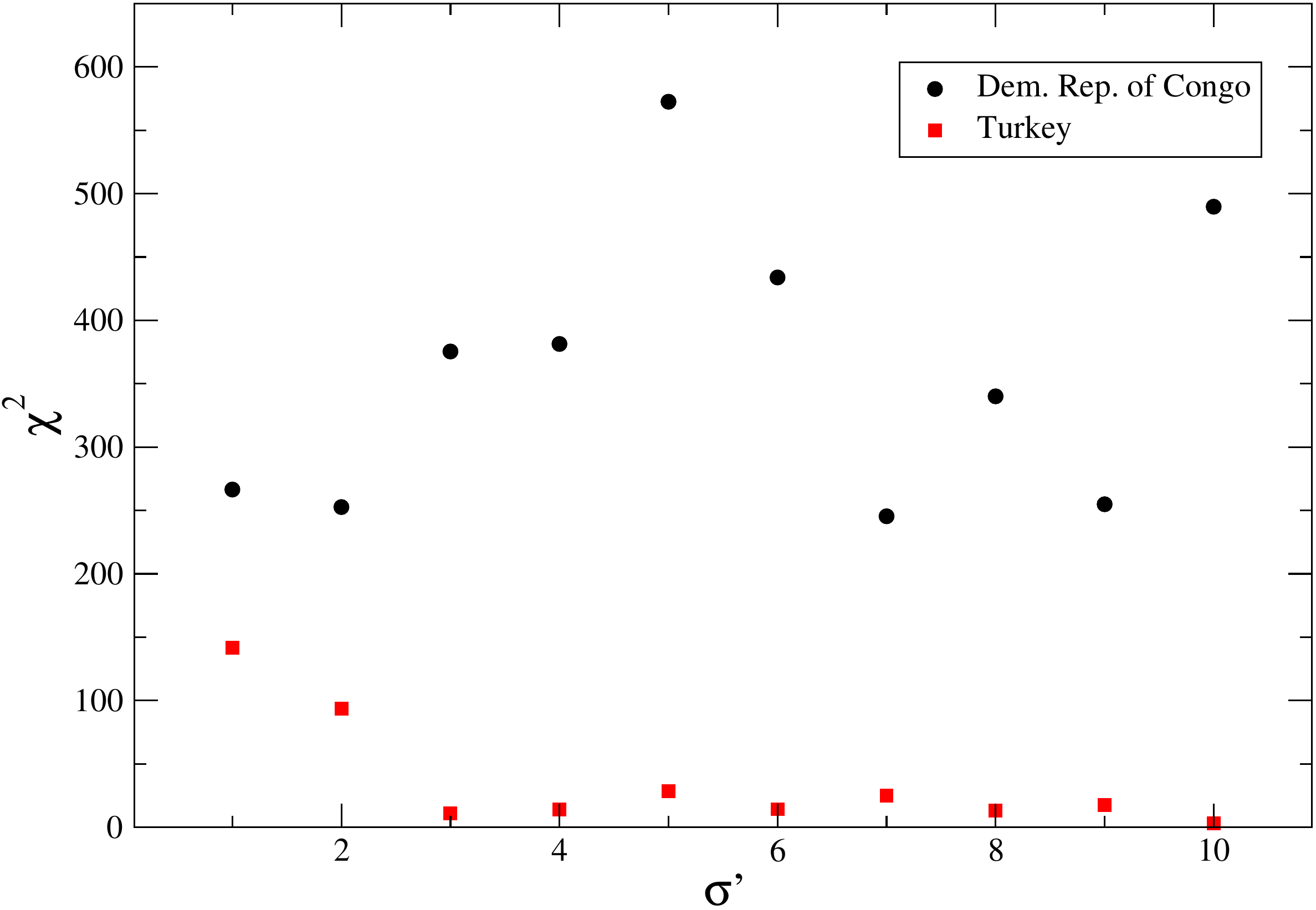}}
\end{center}
        \caption{$\chi^2$ statistic for different values of the rescaling parameters $\sigma^\prime$ for
	the Democratic Republic of Congo and Turkey. The value retained for $\chi^2$ is $245.452$ for the
	former and $3.333$ for the latter.\label{sigmaex}}
\end{figure}

In the present analysis, all data will be rescaled according to the above procedure, but
considering the smallest value for the $\chi^2$ statistic by varying $\sigma^\prime$ in the interval from
$1$ to $10$. Higher values of $\sigma^\prime$ tend to flatten out any discrepancies in the data into the queue of the distribution.
Figure~\ref{sigmaex} shows the different values of $\chi^2$ obtained for the Democratic Republic of Congo and Turkey
as an illustration of this procedure. The values retained in our analysis are then the smallest of those displayed in the Figure.
This will enable, as explained here, a more fair comparison among different countries and localities, and avoiding
frequent false negative results as reported in previous works.

\section{Data sources}
\label{datasourcesec}

All data encompasses the period from the start of the pandemic in each locality (first detected case) up to March, 29 2022.
The following data sources for COVID-19 cases and deaths were used in the present work:
\begin{itemize}
        \item Countries: World Health Organization Coronavirus (COVID-19) Dashboard.
                Data available at {\it https://covid19.who.int/WHO-COVID-19-global-data.csv}.
        \item New York Times COVID-19 cases and deaths tracker. Data available at
                {\it https://raw.githubusercontent.com/nytimes/covid-19-data/master/us-counties.csv}
        \item Brazilian Health Ministry. Data available at {\it https://covid.saude.gov.br/}.
	\item Additional data on number of tests per million inhabitants are available at\\
		{\it https://www.worldometers.info/coronavirus/\#countries}
	\item Excess deaths by COVID-19. Tracking COVID-19 excess deaths across countries. The Economist~\cite{economist}.
\end{itemize}

The economic and social indices used in our analysis are:
\begin{itemize}
        \item HRV Transparency Project: data from 2013. An index based on the public availability of credible
                aggregate economic data~\cite{hollyer}.
        \item Varieties of Democracy Project (V-Dem):  from 2021. Five indices considering the complexity of the concept of democracy:
                Electoral democracy, Liberal democracy, Participatory democracy, Deliberative democracy and
                Egalitarian democracy~\cite{vdem}.
        \item Transparency International Corruption Perception Index: data from 2021.
                A measure of the perceived level of corruption in each country~\cite{transparency}.
        \item Gross Domestic Product per capita (nominal) and by Purchasing Power Parity (PPP): Data from 2017~\cite{gdp}.
        \item Gini index: Data from 2021. An index that measures the degree of income distribution in the population
                of a given country~\cite{gini}.
        \item Literacy rate: data from 2017. UNESCO index of proportion of literacy among adults~\cite{literacy}.
        \item Life expectancy: data from 2021. Life expectancy at birth for each country~\cite{gini}.
        \item Health expenditure (total and per capita): data from 2019. Total and per capita health expenditure
                as a fraction of GDP~\cite{gini}.
        \item Total deaths and cases per million inhabitants: data from April, 2022.
\end{itemize}

\section{Results}
\label{secresults}

\subsection{World}

\begin{figure}[ht]
	\begin{center}
        \hspace*{-6mm}{\includegraphics[width=170mm]{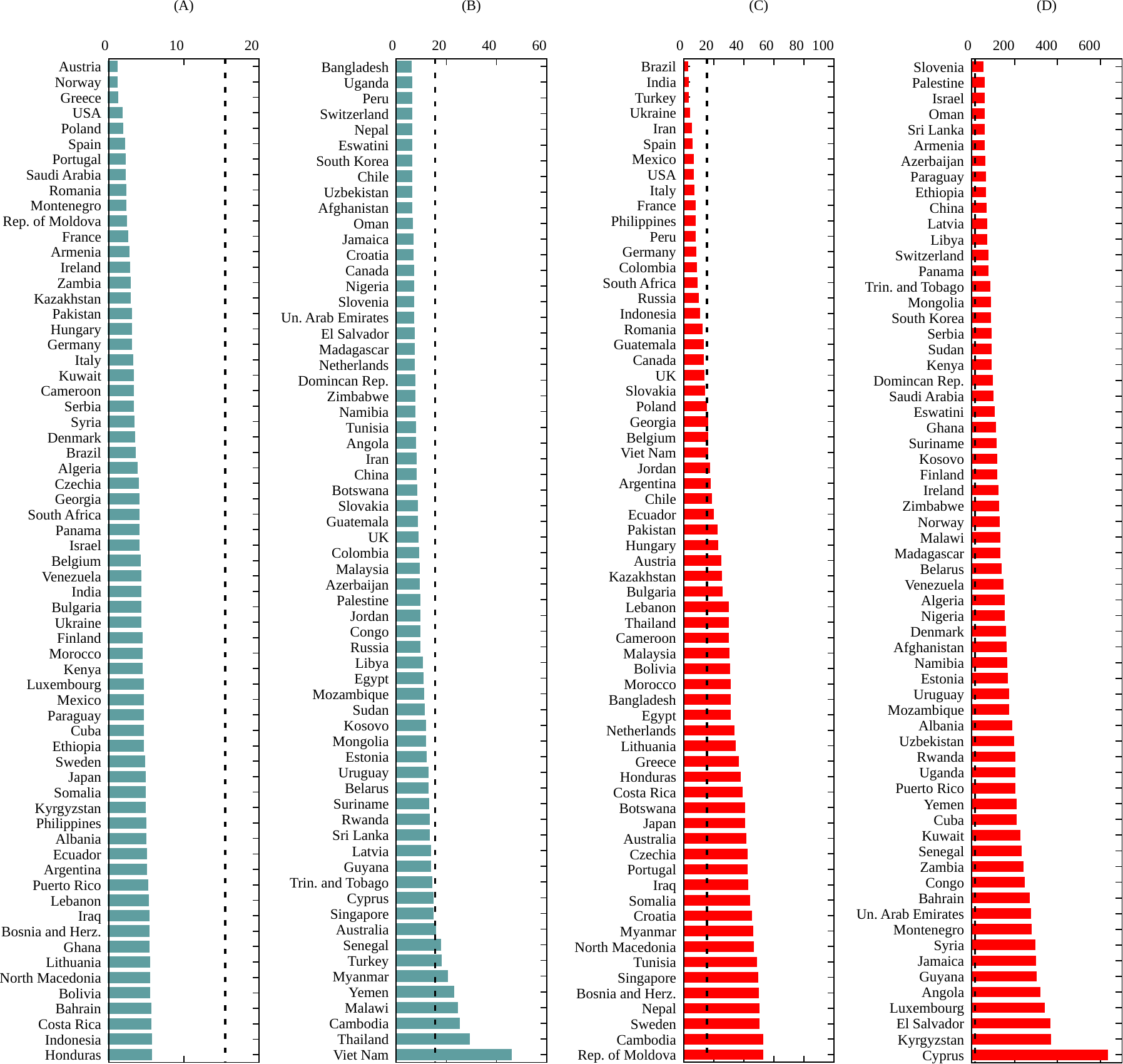}}
	\end{center}
	\caption{$\chi^2$ statistics deviation for each country (with at least 1000 deaths by April, 29 2922)
	for the daily number of new cases (A) and (B), and for daily deaths (C) and (D).
	The dashed line introduced as a reference represents the critical value $\chi^2=15.507$ for rejecting
	the null hypothesis that the data do not conform to the NB distribution.
	\label{figNBworld}}
\end{figure}

We consider data for daily cases and deaths by COVID-19 in the 129 countries with available data in the WHO dashboard,
and with at least one thousand deaths by April, 29 2022. We then determine the first significant digit frequency
for each country, and compute the corresponding $\chi^2$ statistics following the procedure in Sec.~\ref{secrevisit}.
Results are shown in Fig.~\ref{figNBworld} for the daily number of cases, with only 8 countries with a $\chi^2$ value greater than the critical value
for rejecting the null hypothesis. For daily deaths the majority of countries (106 out of 129)
do not comply with the NB law. Considering that the number of deaths is essentially a proportion of the number of cases
(the case fatality ration), this discrepancy points to possible problems in accounting and publicizing the number of deaths
by COVID-19, that requires a closer analysis.

We compute the Spearman rank order correlation between the $\chi^2$ statistic value
for all countries with the economic and social indices listed in Sec.~\ref{datasourcesec}.
The Spearman correlation is given by the Pearson
correlation between the orderings of the values of each data point in the respective sets~\cite{dodge},
and is a measure of the tendency of one variable to be a monotonic function of the other variable.
It's value ranges from $+1$ for an increasing function to $-1$ for a decreasing function.
The results are given in Table~\ref{figNBworld}, and shown some non-negligible correlations that can be used to interpret our results
(see Section~\ref{secdiss}).

\begin{table}[ht]
	\begin{center}
\begin{tabular}{lccc}
        \hline\hline
	Index & $\chi^2$ daily cases & $\chi^2$ daily deaths & N${}^o$ of Countries\\
        \hline\hline
	HRV Transparency & -0.387 & -0.510 & 95 \\
	Corruption perception & -0.202 & -0.125 & 126 \\
	Electoral democracy & -0.203 & -0.198 & 125 \\
	Liberal democracy & -0.237 & -0.213 & 125 \\
	Participatory democracy & -0.241 & -0.259 & 125 \\
	Deliberative democracy & -0.226 & -0.198 & 125 \\
	Egalitarian democracy & -0.252 & -0.190 & 125 \\
	GDP total nominal & -0.304 & -0.584 & 122 \\
	GDP per capita (PPP) & -0.310 & -0.104 & 122 \\
	GDP per capita (nominal) & -0.211 & -0.213 & 122 \\
	GINI & 0.175 & -0.086 & 70 \\
	Literacy rate & -0.233 & -0.141 & 39 \\
	Life expectancy & -0.250 & -0.239 & 127 \\
	Health expenditure (\% of GDP) &  -0.254 & -0.168 & 120 \\
	Health exp.\ per capita (\% of GDP) & -0.031 & 0.398 & 120 \\
	Health exp.\ per capita (nominal) & -0.324 & -0.201 & 118 \\
	Health exp.\ total (nominal) & -0.319 & -0.584 & 118 \\
	Total cases per Million & -0.264 & -0.176 & 114 \\
	Total deaths per Million & -0.206 & -0.361 & 114 \\
	Tests per Million & -0.267 & -0.127 & 114 \\
	Total population & -0.061 & -0.495 & 128 \\
	Excess deaths & 0.055 & 0.151 & 78 \\
	\hline\hline
\end{tabular}
	\end{center}
	\caption{Spearman correlation index between $\chi^2$ from the NB law for each country and
	different social indices. The number of countries in the last column corresponds to the those countries
	considered in our analysis (countries with at least 1000 deaths) and with data available
	for the given index.\label{tabcorr}}
\end{table}

Similarly, we compute the Spearman correlation between each relevant index and the proportion of excess deaths by COVID-19,
as reported in~\cite{economist}. The results are show in Table~\ref{tabcorr2}.

\begin{table}[ht]
        \begin{center}
\begin{tabular}{lc}
        \hline\hline
        Index & $r_s$ \\
        \hline\hline
        HRV Transparency & -0.171 \\
        Corruption perception & -0.558 \\
        Electoral democracy & -0.569 \\
        Liberal democracy & -0.581 \\
        Participatory democracy & -0.572 \\
        Deliberative democracy & -0.581 \\
        Egalitarian democracy & -0.575 \\
        GDP total nominal & -0.363 \\
        GDP per capita (PPP) & -0.480 \\
        GDP per capita (nominal) & -0.333 \\
        GINI & 0.122 \\
        Literacy rate & -0.047 \\
        Life expectancy & -0.541 \\
        Health expenditure (\% of GDP) & -0.395 \\
        Health exp.\ per capita (\% of GDP) & -0.156 \\
        Health exp.\ per capita (nominal) & -0.319 \\
        Health exp.\ total (nominal) & -0.528 \\
        Total cases per Million & -0.470 \\
        Total deaths per Million & -0.208 \\
        Tests per Million & -0.488 \\
        Total population & 0.197 \\
        \hline\hline
\end{tabular}
        \end{center}
        \caption{Spearman correlation index between the proportion of excess deaths for each of the $78$ countries
	with available data and the different social indices.\label{tabcorr2}}
\end{table}

\subsection{United States}

The results for the $\chi^2$ statistic for each state in the USA are shown in Fig.~S1.
Data for daily cases comply with the the NB law for all states. Nevertheless the picture is quite different for
the number of daily deaths, with only 16 states in conformity. We also considered the $1562$ counties in the USA with
at least one hundred deaths. The corresponding histogram for the distribution of values of $\chi^2$ for the daily number of cases and deaths
are shown in Fig.~S2. They are both well fitted by a log-normal distribution, which may evidence
some randomness in the different policies implemented and in the data. Considering the critical value of $\chi^2=15.507$ for
$\alpha=0.05$ only $35\%$ of the counties are in conformity with the NB law for the number of daily cases
and $0.19\%$ for daily deaths.

\subsection{Brazil}

Figure~S3 shows the values of the $\chi^2$ statistic for the 27 Brazilian states and the Federal District.
All states comply with the NB law for the number of cases, but most of then do not comply for the
time series of daily deaths, although, as also occurs for the USA, summing the data for each state
results in a time series of deaths that fully comply with the NB law.
Considering the $858$ municipalities with at least $100$ deaths, $14.5\%$ comply with the NB law for the number of daily cases,
while only two ($0.23\%$) strictly comply for the number of daily deaths. The normalized histogram for the proportion of municipalities
according to the $\chi^2$ statistics are shown in Fig.~S4.

\section{Discussion and conclusions}
\label{secdiss}

We discussed that data drawn from a log-normal distribution with the parameter $\sigma$ greater that one complies with
the Newcomb-Benford first digit distribution, while the converse occurs if $\sigma<1$~\cite{hill}.
In the former case there is no reason from a practical perspective to identify the data as anomalous or
suspect of some kind of manipulation, as both cases are derived from a purely random distribution naturally occurring in many instances
of natural process. in fact, as pointed out above, most if not all of the examples of data satisfying the NB law have
a log-normal distribution. Therefore it is natural to rescale the data such that the resulting distributions for
the modified data have the same average $\mu$ and standard deviation $\sigma$ for the logarithm of the data,
and then apply the NB law test. In our proposed approach
we considered $\mu=0$ and the smaller value of $\chi^2$ obtained in the interval $1<\sigma<10$.
In this way, a significant deviation from the NB law corresponds to a real anomaly in the data, and thus requiring a more
thorough analysis. On the other hand, if the original data have a log-normal distribution, this procedure
results in a compliance with the NB law. This point was clearly illustrated for the daily number of deaths by COVID-19 data
for the whole World, which has a significant $\chi^2$ deviation from the NB law due to
the small value of $\sigma=0.404$, albeit the data distribution being close to a log-normal distribution.
We showed that the rescaling in the data yielded a a very good compliance with the NB law.

Our approach was then applied to the daily number of cases and deaths in each country with at least one thousand
deaths by April, 29 2022. As shown in Fig.~\ref{figNBworld}, for almost all countries the data for the daily number of cases
comply with the NB law, while the deviation for the daily number of deaths is significantly higher, with almost all countries
not complying with the NB law. This is rather unexpected as the number of deaths results from a
convolution of the distribution of the number of days between the first symptom, which is given by a log-normal
distribution~\cite{verity,transnational} and the convolution of two normal distributions are usually well approximated by another log-normal
distribution~\cite{mitchell}.
In order to identify the sources of such deviations we considered the
publicly available social and economic indices listed in Sec.~\ref{datasourcesec}, and computed the Spearman correlation
of each one with the $\chi^2$ deviation from the NB law for countries, as shown in Table~\ref{tabcorr}.
Overall the results point to the fact that lesser democratic and transparent countries
tend to present anomalies in their COVID-19 data. Such deviations are more pronounced for the daily number of deaths.
One result stands out from Table~\ref{tabcorr}:
the Spearman correlation between the value of the $\chi^2$ statistic for the NB distribution for daily deaths and the per capita
health expenditure as a proportion of its GDP of $r_s=0.398$. This indicates that data anomaly for daily deaths (but not daily cases)
is more frequent for countries that expend a higher portion of its GDP in the health sector.
This seems contradictory at first sight, but is explained by the fact that the
per capita nominal expenditure on health services is negatively correlated with the $\chi^2$ deviations for both
daily cases and deaths. This points to the fact that countries with a smaller GDP but that expends a reasonable proportion
of it on health services tend to manipulate the death toll, possibly in order to counter the effects of a failed policy
despite the amount of money spent. In fact, the negative correlation of the daily number
of deaths with the $\chi^2$ deviation shows that data manipulation occurs to hide the real death toll, and this is
more frequent in less democratic and less transparent countries. Also, as expected, the higher the number of tests per million inhabitants,
the lesser the anomalies detected in the data.
The fact that anomalies are much more frequent in the data series for deaths
is probably partially explained by the fact that the social burden of deaths is much more intense than the nominal
number of people manifesting a disease.

\begin{figure}[ht]
        \begin{center}
                {\includegraphics[width=150mm]{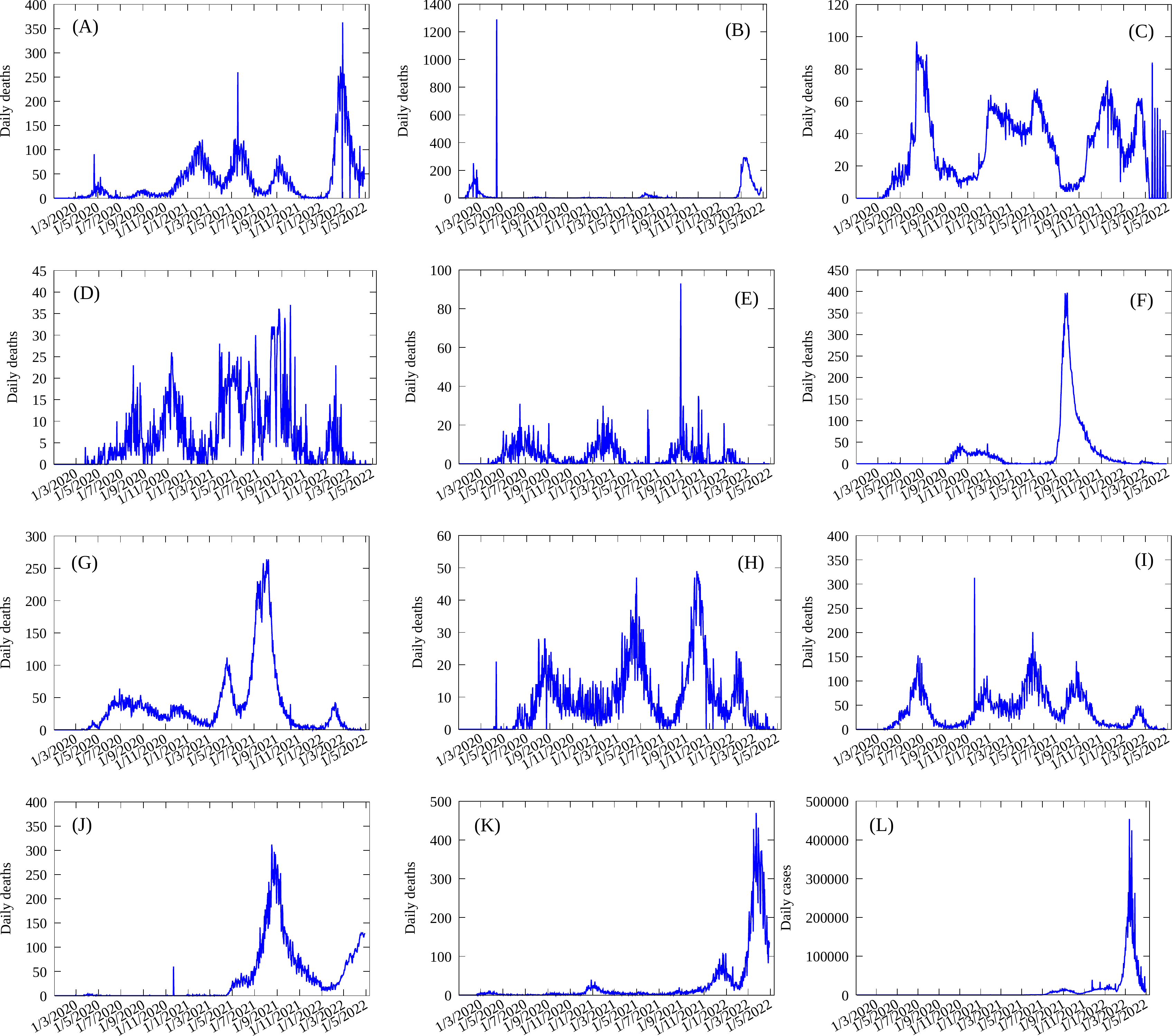}}
       \end{center}
	\caption{Daily deaths by COVID-19 in (A) Japan, (B) China, (C) Egypt, (D) Kenya, (E) Nigeria,
	(F) Myanmar, (G) Bangladesh, (H) Ethiopia, (I) Pakistan, (J) Thailand, (K) South Korea and
	(L) daily cases in Viet Nam.
	\label{daily_deaths_countries}}
\end{figure}

Another explanation for the discrepancy between the $\chi^2$ deviation for the daily number of cases and deaths is that the
span in orders of magnitude for the latter is much smaller than for the former,
as the mortality by COVID-19 is usually in the range of $0.3\%$ to $0.7\%$ of
the real number of cases, which is a few times the official publicized number of cases~\cite{rahmandad}. Therefore the daily
number of deaths is roughly two orders of magnitude the daily number of cases. For countries with smaller populations this can
result is a non compliance with the NB law (see for instance~\cite{nigrini} for a discussion on the conditions for applying the
NB law to empirical data). This is reinforced by the significant negative Spearman correlation between the total population in each country
and the $\chi^2$ deviation from the NB law for the daily number of deaths, showing that smaller populations tend to result in a higher
deviation. On the other hand, the Spearman correlation between the population and the daily number of cases is not significant,
as expected due to the higher range in order of magnitudes of the data. Table~\ref{tab2} shows the values of the $\chi^2$
statistics for the 27 countries with more than 50 million inhabitants. In this cases, the statistics of daily number of deaths spans
a broader range of orders of magnitude. Among those countries, 11 have a significant deviation
from the NB law, and we discuss each separately below.
The corresponding data violating the NB law are shown in Fig.~\ref{daily_deaths_countries}.

\begin{table}[ht]
        \begin{center}
\begin{tabular}{lcc}
        \hline\hline
	Country & $\chi^2$ daily cases & $\chi^2$ daily deaths \\
        \hline\hline
            Brazil & 3.45 & 2.62 \\
	    China & 7.79 & 68.4 \\
	    Egypt & 10.4 & 31.1 \\
	    France & 2.43 & 7.51 \\
	    Germany & 3.00 & 8.08 \\
	    India & 4.20 & 2.93 \\
	    Italy & 3.12 & 6.72 \\
	    Japan & 4.76 & 40.7 \\
	    Kenya & 4.42 & 91.3 \\
	    Mexico & 4.53 & 6.31 \\
	    Myanmar & 20.1 & 45.8 \\
	    Nigeria & 6.68 & 153.3 \\
	    Turkey & 17.6 & 3.33 \\
	    Bangladesh & 5.67 & 30.8 \\
	    Colombia & 8.66 & 8.46 \\
	    Ethiopia & 4.56 & 65.2 \\
	    Indonesia & 5.61 & 10.7 \\
	    Pakistan & 2.95 & 22.2 \\
	    Philippines & 4.85 & 7.53 \\
	    South Africa & 3.96 & 8.94 \\
	    Thailand & 29.0 & 29.6 \\
	    Viet Nam & 45.7 & 16.2 \\
	    South Korea & 5.95 & 89.14 \\
	    Russia & 9.32 & 9.92 \\
	    United Kingdom & 8.50 & 13.45 \\
	    Iran & 7.62 & 5.28 \\
	    United States & 1.72 & 6.34\\
        \hline\hline
\end{tabular}
        \end{center}
	\caption{$\chi^2$ deviation from the NB law for the daily number of cases and deaths for the 27 countries
	with more than 50 million inhabitants.\label{tab2}}
\end{table}

\begin{itemize}

\item {\bf Japan}

The case of Japan illustrates well how the present approach is more sensible to anomalies.
The original (non rescaled) data complies with the NB law with $\chi^2=7.23$, but not if rescaled. This
can be explained by looking at the data for the number of daily deaths in Fig.~\ref{daily_deaths_countries}A. We notice
four spikes in the data, probably due to updates in the data for the total number of deaths. By removing those
four points we obtain the value $7.60$ for the $\chi^2$ statistics for the rescaled data,
which then complies withe the NB law.

\item {\bf China}

The value for the $\chi^2$ deviation for China in Table~\ref{tab2} points to a strong deviation from the NB law,
but at variance with Japan, removing the more visible spikes in the data shown in Fig.~\ref{daily_deaths_countries}B
does not result in a significant reduction of the value of $\chi^2$, pointing to more serious problems with the data.
A more thorough analysis is then in demand to understand this deviation and if some type of data manipulation occurred.

\item {\bf Egypt}

More recently the number of deaths were reported only weakly in Egypt, which is detected as a type
of anomaly in the data. By removing this final period we obtain $\chi^2=19.3$, closer to the limit value
but still deviating from the NB law.

\item {\bf Kenya}

The high value for the $\chi^2$ deviation is due to the relatively small total number of $5649$ deaths, despite the
large population in the country. By itself, this is a point to look closely at how data was collected.
Nevertheless, with a maximum value of $37$ deaths in a single day, the data barely spans two orders of magnitude
and is not expected to conform to the NB law~\cite{nigrini}.

\item {\bf Nigeria}

Nigeria has the highest $\chi^2$ deviation for the daily number of deaths among the countries with a large population.
Again, this cam be explained from the small number
of $3143$ total deaths and a maximum of $93$ daily deaths. Nevertheless, we would expect the data distribution to
be closer to the NB law, but a closer examination of the data shown in Fig.~\ref{daily_deaths_countries}E
shows large fluctuations in the data from day to day, which can be due to poor data gathering or some other
unidentified issue.

\item {\bf Myanmar}

The data for both daily cases and deaths are clearly non-conforming with the NB law. With a total of $19434$ deaths
and a maximum of $397$ daily deaths, there is no possible explanation for such deviation, and a closer analysis
is also in demand in this case. It is important to notice the recent political turmoil in the country, with a military
coup in 2021, is certainly not kin to transparency in public policies.

\item {\bf Bangladesh}

The total number of deaths is Bangladesh is $29127$ and the maximum number of deaths in a day is $264$. Therefore
the high $\chi^2$ deviation points to some type of unexplained data anomaly.

\item {\bf Ethiopia}

For Ethiopia, the small number of deaths, with a maximum number of $49$ daily deaths, makes it difficult to use
the deviation from the NB to pinpoint any anomalies in the data.

\item {\bf Pakistan}

Even removing the single data spike in the data (see Fig.~\ref{daily_deaths_countries}I) the value
of $\chi^2$ does in fact get higher with $\chi^2=25.9$. The noise in the data is unexpectedly significant
for a total of $30369$ deaths and a maximum of $313$ deaths per day. The death toll up to the end of the
considered time window is $0.014\%$ and should be compared to that of Brazil ($0.3$) that has approximately
the same population, and to most western European countries ($\approx 0.2\%$), that have a much better public health
infrastructure. It becomes clear that the data for the number of deaths is not only greatly underestimated,
but also shows evidence of non-explainable data anomalies.

\item {\bf Thailand}

The deviation $\chi^2$ for Thailand is significant for both the number of daily cases and deaths, with
$28400$ total deaths, a maximum number of $312$ deaths per day, $4238061$ total cases and a maximum of
$28379$ reported daily cases. The data should then be expected to comply with the NB law. This non-conformity for both series
of data points to possible data manipulation. This is straighten by the fact that Thailand
has suffered from political instability for a long period of time, and considered to be a flawed democracy~\cite{flawed},
with frequent coups and counting up to 20 constitutions up to the present.

\item{\bf South Korea}

South Korea implemented a tight control of the pandemic during its first year, with a relatively
small incidence of cases and deaths. The relaxation of the most stringent policies led to an important wave
by the end of 2021 and the beginning of 2022. As a consequence, data is biased towards smaller values of daily deaths,
and thence the higher value for $\chi^2$.

\item{\bf Viet Nam}

Viet Nam has the highest value for $\chi^2$ for the daily number of cases relative to the NB law in the World,
despite the fact that number of daily deaths comply with it. The country suffered a very severe wave of COVID-19
starting in February 2022, with a maximum number of $454\,212$ daily cases. The number of daily deaths in the same
period barely exceeded $100$ at least in part due to a high vaccination rate. The present analysis points clearly to some kind
of manipulation of the data for the number of cases.

\end{itemize}

We also performed a closer analysis for the states and cities of the two most attacked countries in the World: the United States
and Brazil, with results shown in Figs.~S1 and~S3.
The daily data for both countries as a whole comply with the NB law in Fig.~\ref{figNBworld}. At the state level,
for the daily number of deaths, the most populous states comply with while those with a smaller population tend to deviate
from the NB law. This comes from the fact that the span in orders magnitude of
the the data is smaller for smaller populations.
A similar picture emerges when looking at the data for
cities (counties in the USA and municipalities in Brazil) with at least $100$ deaths. We observe a large
span for the values of the $\chi^2$ statistic, with a log-normal distribution for both countries,
as shown in Figs.~S2 and~S4, evidencing random fluctuations due to the smaller span of data values.
Nevertheless, this results in country aggregated data satisfying the NB law, and into data closer
to a log-normal distribution, and no evidence of explicit data manipulation.

The proportion of excess deaths can also be used as a metric for the data quality in each country where
this information is available. Unfortunately only 78 countries provide enough information to have
a trusty estimate~\cite{economist}. We computed the Spearman correlation
between the same social and economic indices listed in Sec.~\ref{datasourcesec}, as show in Table.~\ref{tabcorr2}.
We observe a clear tendency for a higher proportion of excess deaths in less democratic, less transparent and more corrupt countries.
Intriguingly, a higher health expenditure and a higher GDP (total and per capita) also imply a higher proportion of non accounted deaths.
On the other hand, and not surprisingly, countries which test less tend to have a higher proportion of deaths in excess by COVID-19.

The present work pointed to different issues in COVID-19 data quality, from possible plain data manipulation, to unaccounted data,
using the NB law and the proportion of excess deaths.
We proposed an improved approach in how to use of the Newcomb-Benford law, but its used is restricted to
data that encompasses at lest a few orders of magnitude, which is the cases for the number of daily cases, but limits its
use to countries (or regions) with a higher population, as shown by the value of the Spearman correlation between the total
population and the $\chi^2$ deviation from the NB law in Table~\ref{tabcorr}. This is an important point to consider when
fitting models using available data.

\section*{Acknowledgments}

This work received financial support from the National Council of Technological and Scientific Development - CNPq (grant
numbers 312857/2021-7 TMRF and 305291/2018-1 MAM).
This work was developed within the scope of the project I3N, UIDB/50025/2020 and UIDP/50025/2020, financed by national funds through the FCT/MEC
(JFFM).

\bibliographystyle{elsarticle-num}

\end{document}


\begin{frontmatter}

\title{Reliability of COVID-19 data and government policies -- Supplemental material.}

\author{T.~M.~Rocha Filho\corref{cor1}}
\ead{marciano@fis.unb.br}
\address{International Center for Condensed Matter Physics and Instituto de Física,\\ Universidade de Brasília, Brasília -- BRAZIL.}
\author{J.~F.~F.~Mendes}
\address{Departamento de Física and I3N, Universidade de Aveiro, 3880 Aveiro -- Portugal}
\author{M.~L.~Lucio}
\address{Departamento de Gestão de Políticas Públicas/FACE, Universidade de Brasília, Brasília -- BRAZIL.}
\author{M.~A.~Moret}
\address{Centro Universitário SENAI CIMATEC and \\ Universidade do Estado da Bahia, Salvador -- Brazil}

\end{frontmatter}

\begin{figure}[ht]
        \begin{center}
        {\includegraphics[width=170mm, angle =-90]{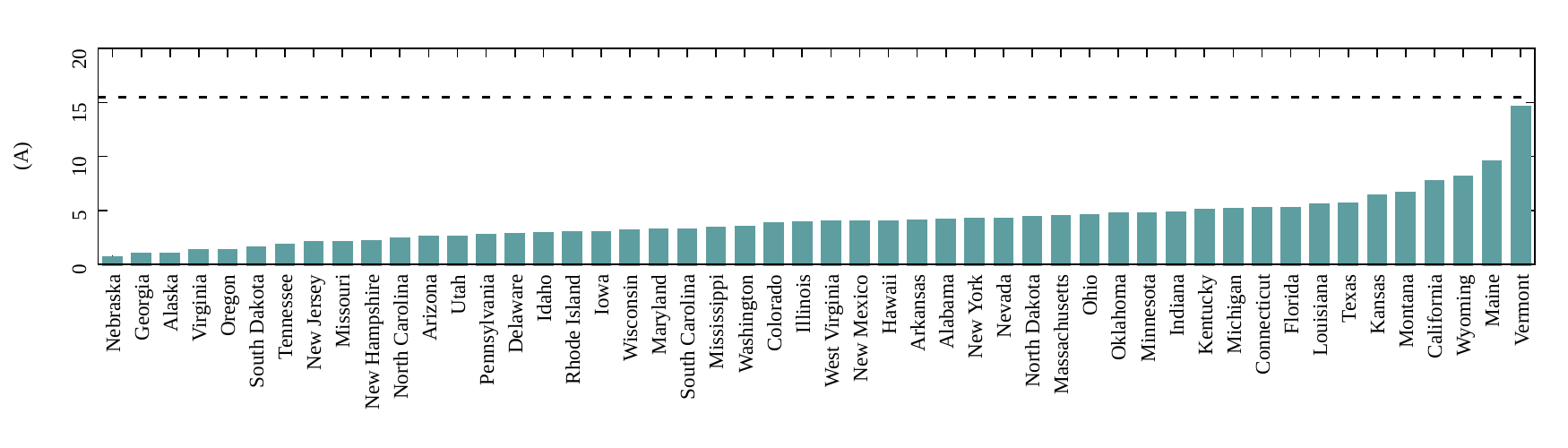}}\hspace*{-3mm}
        {\includegraphics[width=170mm, angle =-90]{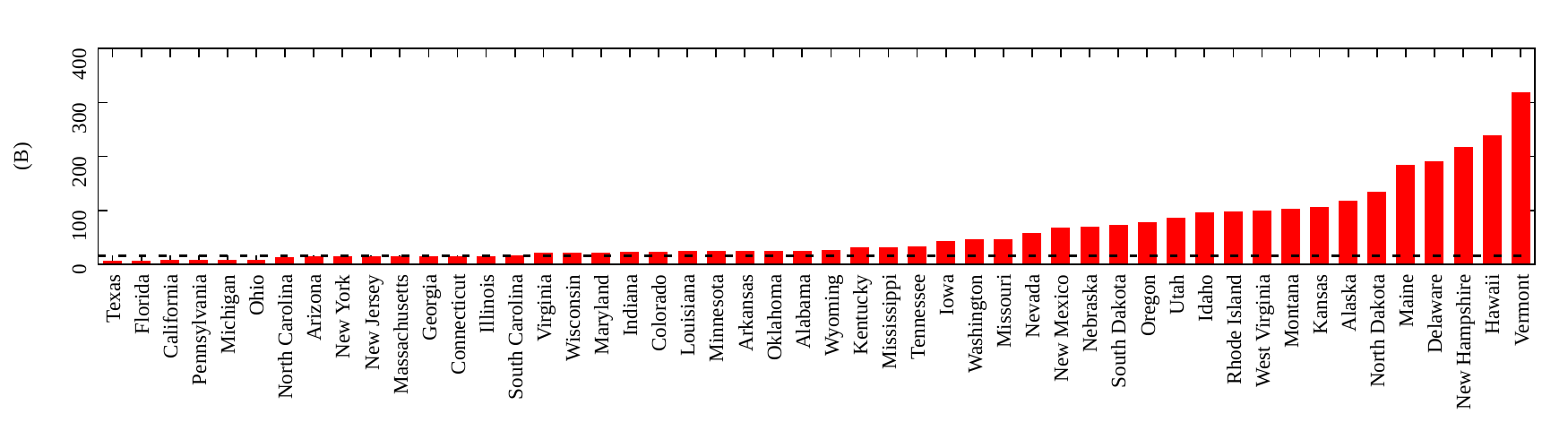}}
        \end{center}
        {Fig.~S1: $\chi^2$ statistics deviation for each USA state for the daily number of new cases (A) and
        and for daily deaths (B).\label{figNB_EUA_states}}
\end{figure}

\begin{figure}[ht]
        \begin{center}
        {\includegraphics[width=140mm]{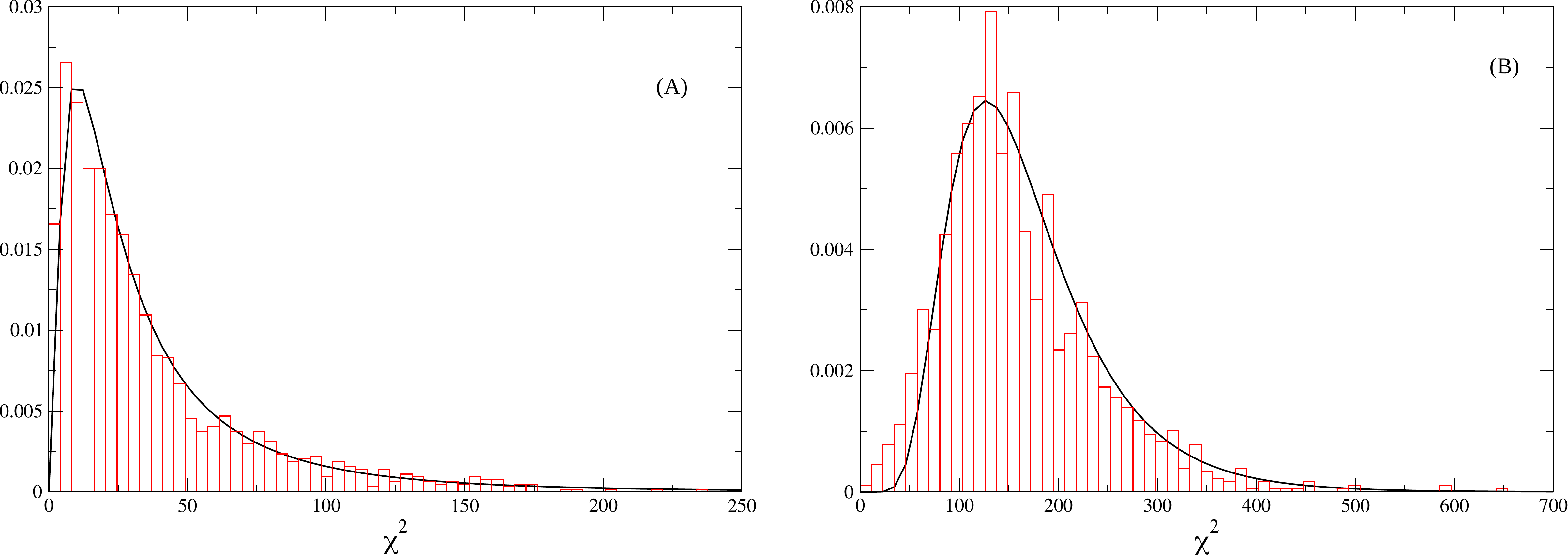}}
        \end{center}
	{Fig.~S2: Normalized distribution of values of the $\chi^2$ deviation from the Newcomb-Benford
	         distribution for USA counties with at least 100 deaths (A) Daily cases and (B) Daily deaths.
		 The black continuous line is the best fit log-normal distribution.
		 \label{distEUA}}
\end{figure}

\begin{figure}[ht]
        \begin{center}
        {\includegraphics[width=170mm, angle =-90]{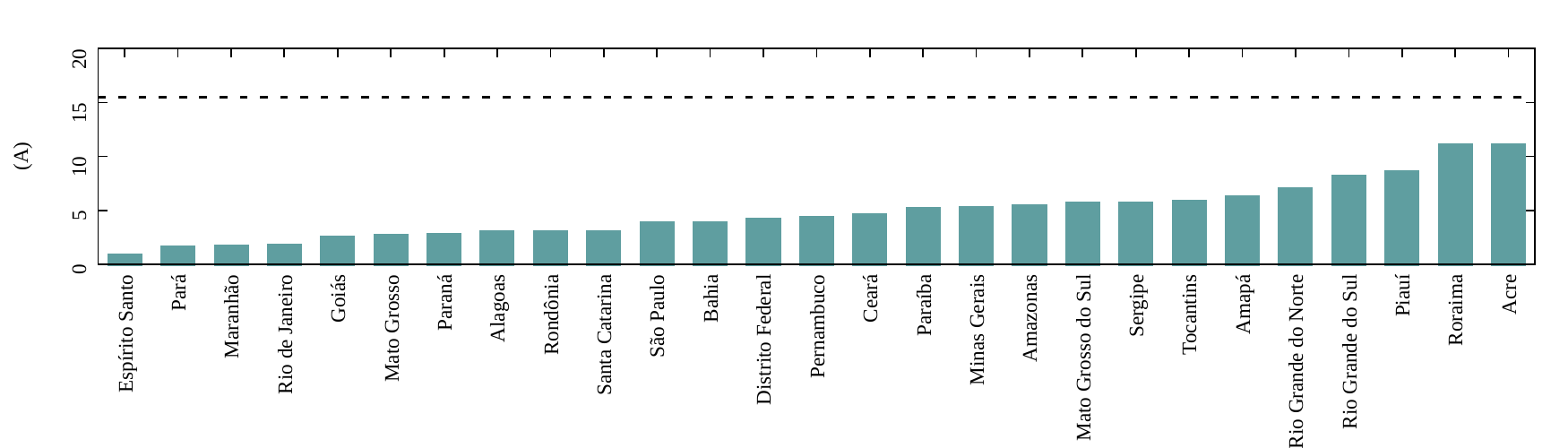}}\hspace*{-3mm}
        {\includegraphics[width=170mm, angle =-90]{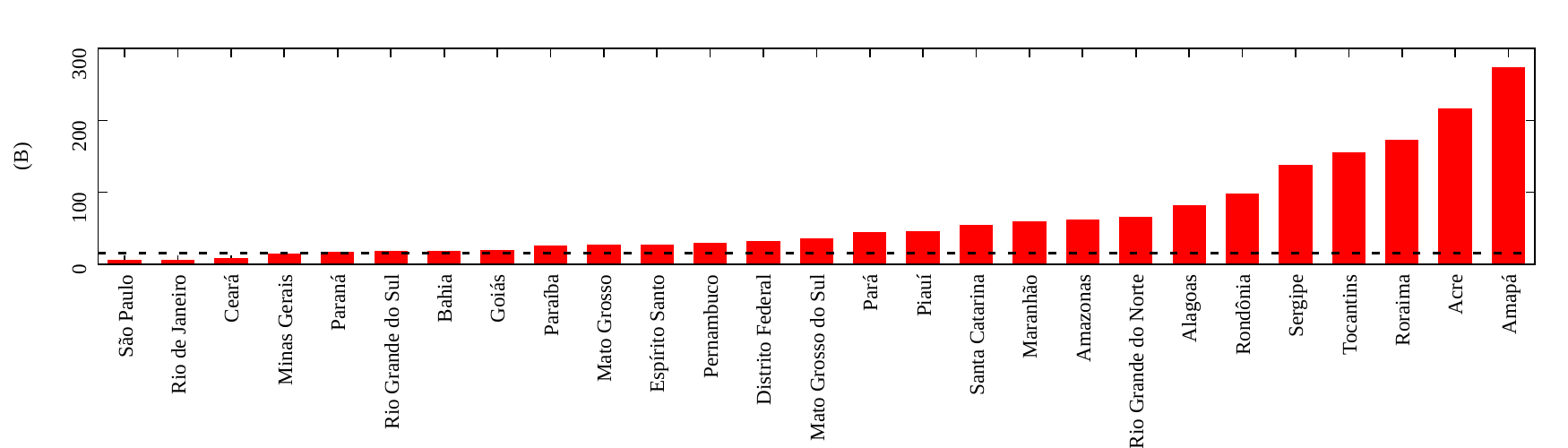}}
        \end{center}
	{Fig.~S3: $\chi^2$ statistics deviation for each Brazilian state for the daily number of new cases (A) and
	and for daily deaths (B)\label{figNB_BR_estados}}
\end{figure}

\begin{figure}[ht]
        \begin{center}
        {\includegraphics[width=140mm]{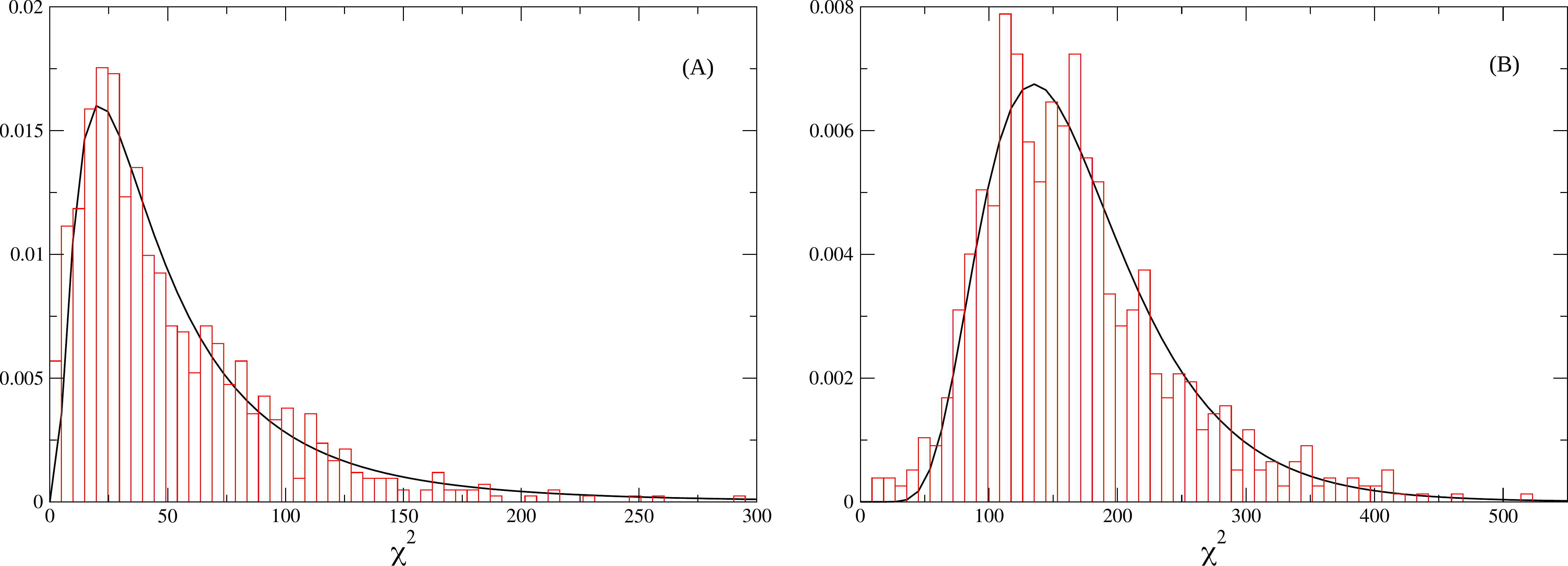}}
        \end{center}
	{Fig.~S4: Normalized distribution of values of the $\chi^2$ deviation from the Newcomb-Benford
		 distribution for Brazilian municipalities with at least 100 deaths (A) Daily cases and
		 (B) Daily deaths. The black continuous line is the best fit log-normal distribution.\label{distBrasil}} 
\end{figure}